\documentclass[twocolumn]{aa}

\newcommand {\apgt} {\ {\raise-.5ex\hbox{$\buildrel>\over\sim$}}\ }
\newcommand {\aplt} {\ {\raise-.5ex\hbox{$\buildrel<\over\sim$}}\ }
\newcommand{\rmn}[1] {\mathrm{#1}}

\usepackage{graphicx}
\usepackage{txfonts}
\usepackage{colordvi}
\def\r[#1]{\Red{#1}}
\begin{document}
   \title{The milliarcsecond--scale jet of PKS\,0735+178 during quiescence}

   \author{I. Agudo
          \inst{1,2}
          \and
          J. L. G\'omez
	  \inst{2}
          \and
          D. C. Gabuzda
	  \inst{3}
          \and
          A. P.  Marscher
	  \inst{4}
          \and
          S. G. Jorstad
	  \inst{4,5}
	  \and
          A. Alberdi
	  \inst{2}
          }

   \offprints{I. Agudo, \email{iagudo@mpifr-bonn.mpg.de}}

   \institute{Max-Planck-Institut f\"ur Radioastronomie,
              Auf dem H\"ugel, 69,
              D-53121, Bonn, Germany
         \and
	      Instituto de Astrof\'{\i}sica de Andaluc\'{\i}a,
	      CSIC, Apartado 3004, 18080 Granada, Spain
         \and
	      Physics Department, University College Cork, Cork,
	      Ireland
         \and
	      Institute for Astrophysical Research, Boston University,
	      725 Commonwealth Avenue, Boston, MA 02215, USA
         \and
	      Sobolev Astronomical Institute, St. Petersburg State
	      University, Universitetskij Pr. 28, 198504, St.
	      Petersburg, Russia\\
             }


   \abstract{

We present polarimetric 5\,GHz to 43\,GHz VLBI observations of
the BL\,Lacertae object PKS\,0735+178, spanning March 1996 to May
2000. Comparison with previous and later observations suggests that
the overall kinematic and structural properties of the jet are greatly
influenced by its activity. Time intervals of enhanced activity, as
reported before 1993 and after 2000 by other studies, are followed
by highly superluminal motion along a rectilinear jet. In contrast
the less active state in which we performed our observations, shows
subluminal or slow superluminal jet features propagating through
a twisted jet with two sharp bends of about 90$^\circ$ within the
innermost three--milliarcsecond jet structure. Proper motion estimates
from the data presented here allow us to constrain the jet viewing angle
to values $\le 9^\circ$, and the bulk Lorentz factor to be between 2 and 4.

   }

   \maketitle
%

\section{Introduction}
\label{intro}

PKS\,0735+178 (0735+178 throughout) was first identified as a
BL\,Lacertae object (BL\,Lac) by Carswell et al.\ (\cite{Car74}).
These authors reported absorption lines redshifted by
$z_{abs} = 0.424$, which was later confirmed by Rector \& Stocke
(\cite{Rec01}). Given that only absorption lines were detected,
this $z_{abs}$ should be considered as a lower limit for the
redshift of 0735+178.
Throughout this paper we assume
$z = z_{abs}=0.424$, a Hubble constant $H_{\circ} =
72$\,km\,s$^{-1}$\,Mpc$^{-1}$
and a Friedmann-Robertson-Walker cosmology with $\Omega_{m}=0.3$
and $\Omega_{\Lambda}=0.7$. Under these assumptions the luminosity
distance of 0735+178 is $d_{L}=2263$\,Mpc, an angular size
of one milliarcsecond (mas) on the plane of the sky corresponds to
a linear size of $5.4$\,pc in the frame of the source, and
an angular motion of 1\,mas/yr corresponds to a speed of $25.1\,c$.

At radio wavelengths, 0735+178 is predominantly point-like at arcsecond
resolution (Ulvestad, Johnston \& Weiler \cite{Ulv83}; Kollgaard
et al.\ \cite{Kol92}). Radio images made from data obtained
with early very long baseline interferometry (VLBI) arrays at
intermediate centimetre wavelengths have typically shown a compact
core and a rather straight jet extending toward the north--east
(e.g., Gabuzda et al.\ \cite{Gab94}).
Multi-epoch VLBI observations made before 1993 allowed the
identification of a number of superluminal jet features, with
apparent speeds in the range $7\,c$ to $12\,c$
(Gabuzda et al.~\cite{Gab94} and references therein).

After 1994, VLBI observations at higher frequencies provided
evidence for a curved structure within the inner jet of 0735+178
(Kellermann et al.\ \cite{Kel98}; G\'omez et al.\ \cite{Gom99}).
In particular, the polarimetric 22\,GHz and 43\,GHz images of G\'omez
et al.\ (\cite{Gom99}), obtained in late 1996, revealed a twisted
jet with two sharp apparent bends within the innermost 3\,mas.
Throughout the rest of the paper, we will refer
to these two bends as the first ($\sim 1$\,mas east of the core,
where the jet turns toward the north)
and second ($\sim2$\,mas north--east of the core, where the jet
turns toward the east)
bends. G\'omez et al.\ (\cite{Gom99}) also found
the polarized emission following the outermost bend in the jet,
with the magnetic vector being parallel to the local jet axis.
Ojha et al.\ (\cite{Ojh04}) showed a
very similar 15\,GHz polarization structure for a nearby epoch.
Based on 8.4\,GHz and 43\,GHz VLBI observations spanning from 1995
to 1998, along with existing previous observations, G\'omez et al.
(\cite{Gom01}) studied the kinematics of the components since the
first VLBI observations of 0735+178, made in 1979. Their
results raised the possibility that the jet proper motions
for epochs after 1994 could be much slower than measured
before 1993.

Comparison of 15\,GHz VLBI observations in February 1999
and 5\,GHz VSOP\footnote{VLBI Space Observatory Program.}
observations in January 1999, both probing the same jet
angular scales, showed markedly different jet structures
at these two frequencies (Gabuzda, G\'omez \& Agudo \cite{GabGA01}).
The most striking difference was the relative weakness
of the 5\,GHz emission near the first bend, which implied a
highly inverted spectrum in this area. This behaviour was shown
to be consistent with free-free absorption of the emission
near this bend.
Evidence for a local enhancement in the Faraday rotation measure
(RM) of ($120\pm55$)\,$\rmn{rad/m^{2}}$ was found for the same region.
This behaviour was interpreted as being consistent with
an interaction of the jet with the external medium surrounding
the region of the first sharp bend, similar to the situation
observed for the VLBI jet of the radio galaxy 3C\,120 (G\'omez et al.
\cite{Gom00}).

Here we present polarimetric 5\,GHz to 43\,GHz VLBI
observations of 0735+178, spanning from March 1996
to May 2000.
This allows us to trace the magnetic field pattern
where the jet executes the two bends. We also carry
out a more complete study of proper motions and
changes in intensity along the jet than has been
possible in earlier studies.


\section{Observations and data reduction}
\label{obs}

The new polarimetric observations presented in this paper were
obtained within two observing programs, both using
the Very Long Baseline Array (VLBA) of the National Radio
Astronomy Observatory (NRAO\footnote{The NRAO of the USA is
operated by Associated Universities, Inc., under co-operative
agreement with the National Science Foundation.}).
The first consisted of four observing sessions of 12\,h performed
on 26 March 1996, 2 October 1996, 6 April 1997 and 18 October
1997 at 8.4\,GHz and 22\,GHz. During the 6 April 1997 observations,
5\,GHz and 15\,GHz were also measured, see Table~\ref{datlst}.
The second observing program consisted of three observing sessions
of 12\,hr at 15\,GHz, 22\,GHz and 43\,GHz performed on 27 February 1999,
1 September 1999 and 20 May 2000. The 15\,GHz image from February
1999 was used for spectral index and RM analysis by Gabuzda,
G\'omez \& Agudo (\cite{GabGA01}).

The total--integration times on 0735+178 were about
$40$\,min at 5\,GHz, $60$\,min at 8.4\,GHz,
$75$\,min at 15\,GHz, $95$\,min at 22\,GHz and
$230$\,min at 43\,GHz.

Both left- and right-circular polarization data were recorded at
each of the VLBA telescopes at 128\,Mbps, with 1\,bit sampling and
eight IFs each of 8\,MHz bandwidth.
The data were correlated at the VLBA correlator
in Socorro, New Mexico. The calibration of the data was performed
with the AIPS package (Fomalont \cite{Fom81}) following the
standard procedure for polarimetric observations (e.g. Lepp\"anen,
Zensus \& Diamond \cite{Lep95}).
Opacity corrections for the 22\,GHz and 43\,GHz data were performed
using the variation of the system temperature with elevation and
solving for the receiver temperature and zenith opacity at each
of the VLBA antennas.
Opacity corrections for the 15\,GHz data were negligible.
The instrumental polarization terms were determined using the
parallactic angle dependence to separate the source polarization
and the instrumental polarization through the feed--solution
algorithm developed by Lepp\"anen, Zensus \& Diamond~(\cite{Lep95}).

After the initial reduction, the data were edited, self--calibrated,
imaged and deconvolved using CLEAN both in total intensity ({\it I})
and polarization ({\it P}) using a combination of AIPS and the
Caltech DIFMAP software (Pearson et al.\ \cite{Pea94}).

Calibration of the absolute orientation of the polarization position
angles ($\chi$) was performed by comparing the $\chi$ values corresponding
to the total polarization in the VLBA images for 0735+178 and various
calibrators and other target sources (PKS\,0420-014, PKS\,0528+134,
OJ\,287, 3C\,279, 3C\,454.3 and BL\,Lac) with those measured with
the VLA at contemporaneous epochs. We also used the data from the
VLA/VLBA Polarization Calibration
service\footnote{http://www.aoc.nrao.edu/$\sim$smyers/calibration/}
and data from the 4.8\,GHz. 8\,GHz and 14.5\,GHz
UMRAO\footnote{The University of Michigan Radio Astronomy
Observatory is supported by funds provided by
the University of Michigan.}
monitoring program as a cross--check. The typical uncertainties in
the $\chi$ values range between $3^{\circ}$ and $5^{\circ}$ at all
frequencies below 43\,GHz; at the single epoch for which we detected
polarized emission in 0735+178 at 43\,GHz (May 2000), the uncertainty
was larger, $\sim 7^{\circ}$.

\begin{table}
\centering
\caption[]{Observing epochs and frequencies.}
\begin{flushleft}
\begin{tabular} {lcc}
\hline\noalign{\smallskip}
Epoch & Date & Freq. (GHz)\\
\hline\noalign{\smallskip}
1996.23 & 26 March 1996    & 8.4, 22\\
1996.75 & 2 October 1996   & 8.4, 22\\
1997.26 & 6 April 1997     & 5, 8.4, 15, 22\\
1997.80 & 18 October 1997  & 8.4, 22\\
1999.16 & 27 February 1999 & 15, 22, 43\\
1999.67 & 1 September 1999 & 15, 22, 43\\
2000.39 & 20 May 2000      & 15, 22, 43\\
\noalign{\smallskip}
\hline
\end{tabular}
\end{flushleft}
\label{datlst}
\end{table}


\section{Results}
\label{res}

\begin{figure}
\centering
\includegraphics[bb=0 0 286 275,width=9cm,clip]{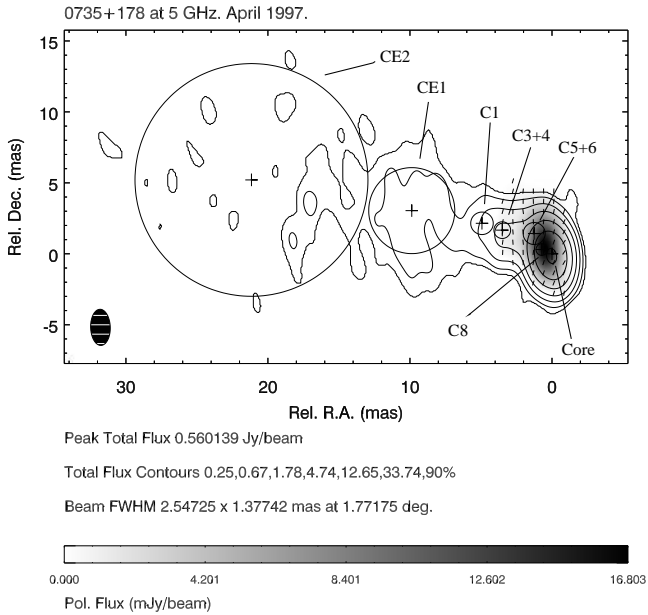}
\caption{5\,GHz VLBA image of 0735+178 on 6 April 1997. The contours
         represent the observed total intensity, the grey scale the
         polarized intensity and the superimposed sticks show
	 the orientation of the polarization electric vectors.
	 The positions of the fitted
	 Gaussian components are indicated by the crosses, whereas the
	 circles (of radius equal to the FWHM of each Gaussian) symbolize
	 their size.}
\label{5ghz}
\end{figure}

\begin{figure}
\centering
\includegraphics[bb=14 14 536 857,width=9cm,clip]{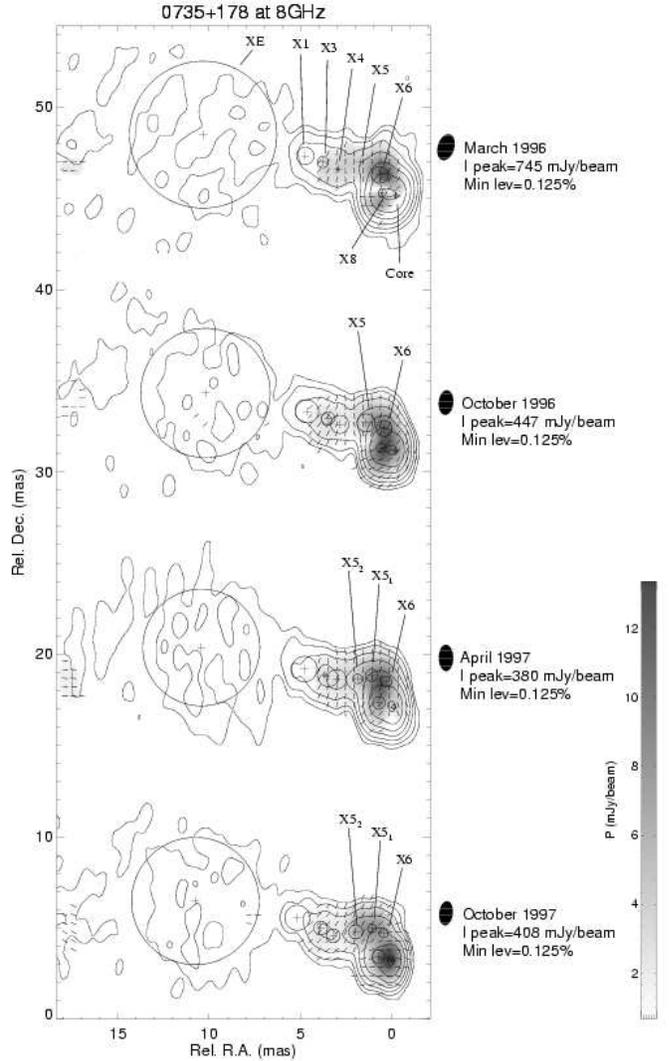}
\caption{8.4\,GHz VLBA images of 0735+178. Observing epochs
         are given for each image. The total intensity is plotted
	 as contours at seven logarithmically equispaced levels,
         starting at the minimum (indicated to the right of each
         image) up to the 90\,\% of the corresponding {\it I} peak
         (also given for each map).  The colour scale shows the
	 distribution of the polarized intensity and the superimposed
	 sticks show the orientation of the polarization electric vectors.
         From top to bottom, the convolving beams (shown as filled ellipses)
         are ($1.46\times0.90$)\,$\rm{mas}^{2}$,
	 ($1.35\times0.80$)\,$\rm{mas}^{2}$,
	 ($1.44\times0.79$)\,$\rm{mas}^{2}$
         and ($1.33\times0.74$)\,$\rm{mas}^{2}$ with major--axis position angles
         at $-16.6^{\circ}$, $-5.4^{\circ}$, $0.6^{\circ}$ and $-6.5^{\circ}$,
         respectively. The positions of the fitted
	 Gaussian components are indicated by the crosses, whereas the
	 circles (of radius equal to the FWHM of each Gaussian) symbolize
	 their size.}
\label{8ghz}
\end{figure}

\begin{figure}
\centering
\includegraphics[bb=14 14 494 857,width=9cm,clip]{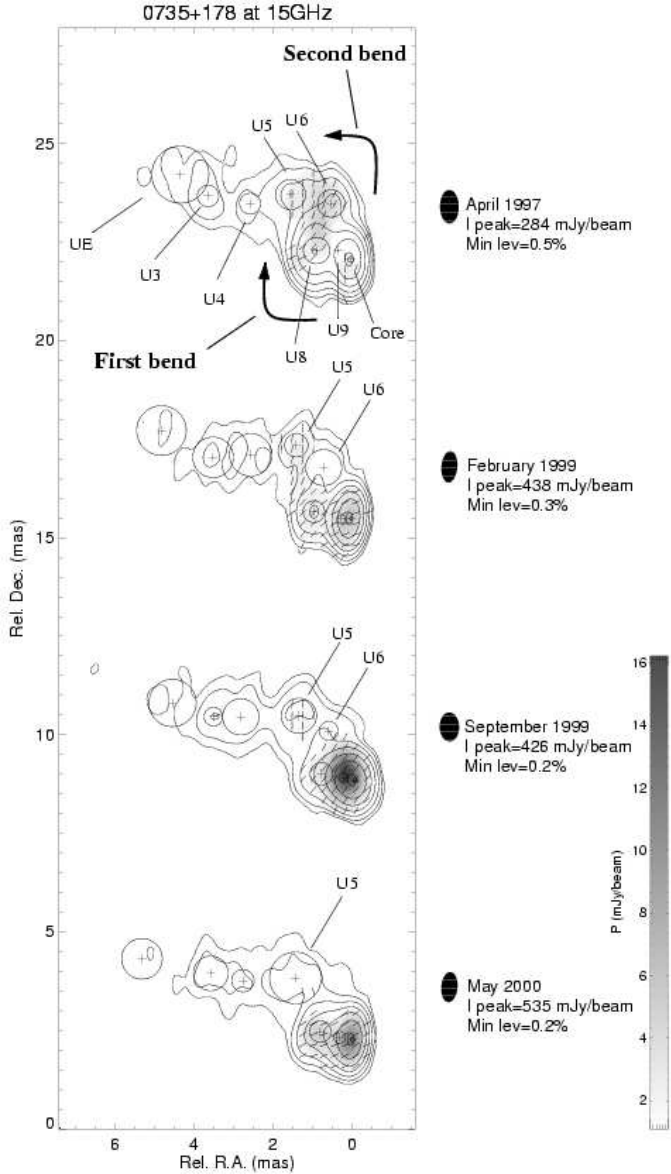}
\caption{Same as Fig. \ref{8ghz} for the 15\,GHz images.
          The location of the first and second bends defined
	 in \S~\ref{intro} are indicated in the image taken
	 on April 1997.
         From top to bottom, the convolving beams are
	 ($0.85\times0.47$)\,$\rm{mas}^{2}$,
         ($0.80\times0.40$)\,$\rm{mas}^{2}$,
	 ($0.73\times0.49$)\,$\rm{mas}^{2}$ and
	 ($0.76\times0.41$)\,$\rm{mas}^{2}$
         with major axis position angles at $0.5^{\circ}$, $-4.7^{\circ}$,
         $1.1^{\circ}$ and $-4.8^{\circ}$, respectively.}
\label{15ghz}
\end{figure}

\begin{figure}
\centering
\includegraphics[bb=14 14 465 833,width=9cm,clip]{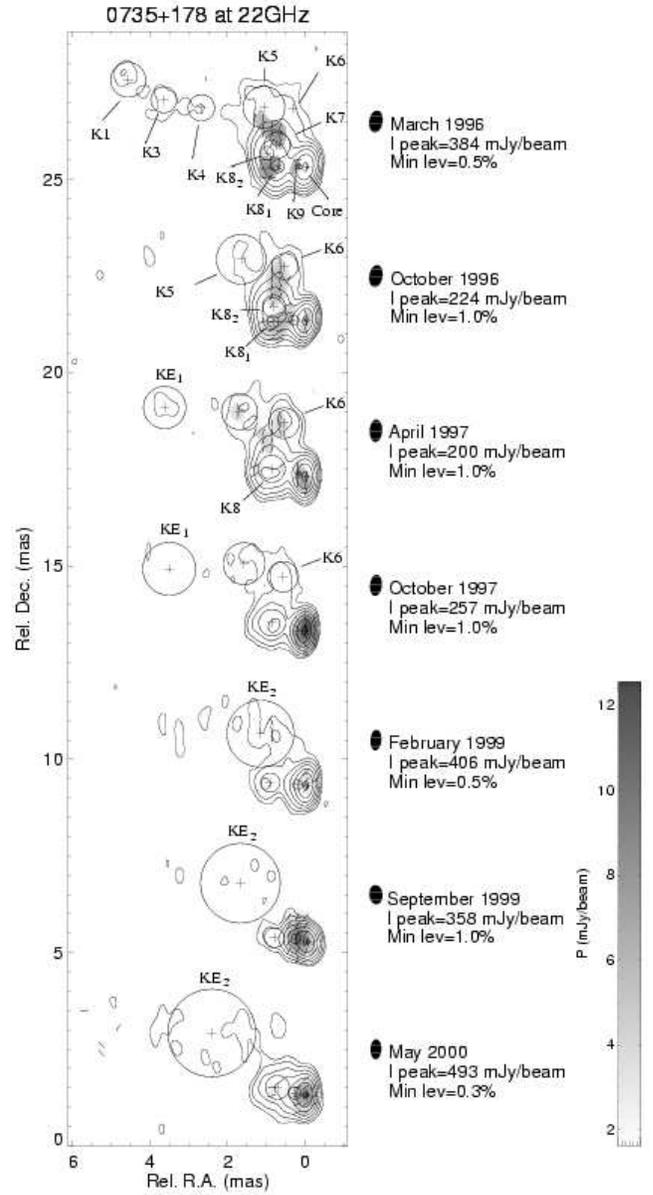}
\caption{Same as Fig. \ref{8ghz} but for the 22\,GHz images. From
         top to bottom, the convolving beams are
	 ($0.57\times0.34$)\,$\rm{mas}^{2}$,
         ($0.57\times0.34$)\,$\rm{mas}^{2}$,
	 ($0.58\times0.32$)\,$\rm{mas}^{2}$,
	 ($0.54\times0.30$)\,$\rm{mas}^{2}$,
         ($0.54\times0.28$)\,$\rm{mas}^{2}$,
	 ($0.49\times0.34$)\,$\rm{mas}^{2}$ and
	 ($0.51\times0.28$)\,$\rm{mas}^{2}$
         with major axis position angles at $-9.5^{\circ}$, $-9.6^{\circ}$,
         $-3.0^{\circ}$, $-6.0^{\circ}$, $-4.7^{\circ}$, $5.4^{\circ}$ and
         $-2.0^{\circ}$, respectively.}
\label{22ghz}
\end{figure}

\begin{figure}
\centering
\includegraphics[bb=0 0 587 840,width=9cm,clip]{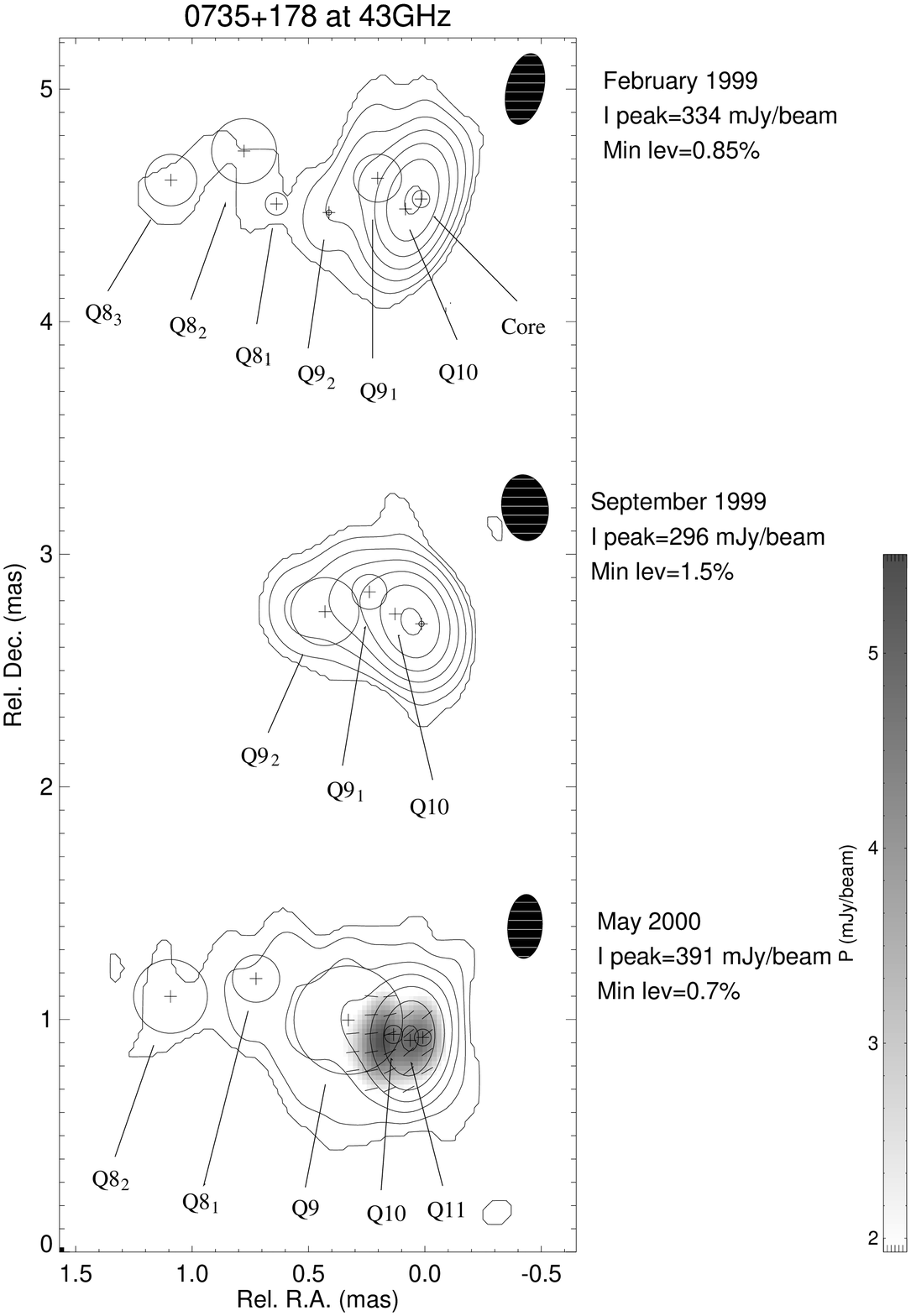}
\caption{Same as Fig. \ref{8ghz} but for the 43\,GHz images. From
         top to bottom, the convolving beams are
	 ($0.31\times0.16$)\,$\rm{mas}^{2}$,
         ($0.29\times0.20$)\,$\rm{mas}^{2}$ and
	 ($0.28\times0.15$)\,$\rm{mas}^{2}$ with major axis
         position angles at $-12.8^{\circ}$, $7.8^{\circ}$ and
         $-3.4^{\circ}$, respectively.}
\label{43ghz}
\end{figure}

The resulting {\it I} and {\it P} images at 5\,GHz, 8.4\,GHz, 15\,GHz,
22\,GHz and 43\,GHz are presented in Figs. \ref{5ghz}-\ref{43ghz}.
The {\it I} images were obtained using uniformly weighted
($u$,$v$)--coverage, whereas the {\it P} (Stokes {\it Q} and {\it U})
images were obtained using natural weighting.

The frequency coverage of our observations allowed us to study
each region of the jet with adequate resolution within the
inner 10\,mas. The bends in the inner jet are not clearly visible
at 5\,GHz (Fig. \ref{5ghz}) due to the relatively low
angular resolution and the weakness of the emission near the first
sharp bend at this frequency (due to low-frequency absorption,
as reported by Gabuzda, G\'omez \& Agudo \cite{GabGA01}).
The VLBI structure at 5\,GHz is dominated by a
compact core plus a broad jet extended toward the north-east up
to $\sim20$\,mas to $\sim30$\,mas from the core.
At the other frequencies, the two
previously reported sharp bends of $\sim 90^{\circ}$ can be seen
within the inner 3\,mas of the jet (Figs. \ref{8ghz},\ref{15ghz}
and \ref{22ghz}).
Comparison of our new 8.4\,GHz, 15\,GHz and 22\,GHz images
and previous {\it I} images (Kellermann et al.\ \cite{Kel98};
G\'omez et al.\ \cite{Gom99}, \cite{Gom01}; Gabuzda, G\'omez \&
Agudo \cite{GabGA01}) shows that the source has maintained its
double bend structure from 1996 to 2000.
However, in the last three epochs, the second bend seems to
decrease its curvature slightly (see Fig. \ref{15ghz}).

The 5\,GHz and 8.4\,GHz maps (Figs. \ref{5ghz} and \ref{8ghz})
show the polarization structure extending to $\sim 5$\,mas from
the core. As was also reported from previous VLBI observations,
the $\chi$ values outside the first bend are aligned
perpendicularly to the local jet axis (see e.g. Figs. \ref{8ghz}
and \ref{15ghz}).
The region of the first bend, and inwards, displays a high degree of
variation in polarization between our observing epochs (Figs.
\ref{8ghz} and \ref{15ghz}).

\subsection{Jet structure modelling}
\label{modfits}

To describe the total intensity distribution
in the 0735+178 jet, we derived circular Gaussian model fits to
all the data sets, in the visibility plane, using the DIFMAP software.
The resulting parameters of the model fits, as well as the
electric--vector position angle $\chi$ and degree of polarization
($m$), computed as the $\chi$ and $m$ averages in boxes of
$3\times3$ pixels around the position of each component, are
listed in Tables \ref{fits5} to \ref{fits43}.

Note that a slightly different 15\,GHz model fit than that presented
by Gabuzda, G\'omez \& Agudo (\cite{GabGA01}) is presented here
for February 1999 (Table \ref{fits15}). The new model contains
an extra component in the region between 3\,mas and 4\,mas
from the core. The reason for changing this model was to make it
more consistent with the modelling of this region at 15\,GHz for
the remaining observing epochs by considering the same number
of components across epochs.
Note, however, that this does not affect in any way the validity
of the results presented by Gabuzda, G\'omez \& Agudo (\cite{GabGA01}).

To obtain a good representation of the jet structure at
22\,GHz (epochs March 1996 and October 1996) and 43\,GHz (epochs
February 1999, September 1999 and May 2000) it was necessary to
split the emission regions corresponding to components i8 and i9
(with i=C, X, U; see Figs. \ref{5ghz} to \ref{43ghz} and Tables
\ref{fits5} to \ref{fits43}) into several sub-components, labelled
K8$_1$, K8$_2$, Q8$_1$, Q8$_2$, Q8$_3$, Q9$_1$ and Q9$_2$.
For a better characterization of all the jet
components it was also necessary to fit the extended emission
observed in the outer regions of the jet (modelled as components
iE with i=C, X, U, K; see Figs. \ref{5ghz} to \ref{22ghz} and Tables
\ref{fits5} to \ref{fits22}). Note that these components
should not be taken as representing real distinct features in the jet,
but instead extended regions with weak flux densities.

The errors in the DIFMAP model-fit parameters
(total flux density, position and size of the components) were obtained
using the DIFWRAP software, developed by Lovell (\cite{Lov00}).
The errors were determined by perturbing the fit parameters
until the resulting residual maps were
unacceptable. These perturbed quantities defined the upper and
lower limits for the errors of the fit parameters.
To take into account the interrelation between the fit parameters,
we also allowed for simultaneous variations of all
four parameters for each component when determining the final
uncertainties.
This resulted in somewhat larger but more conservative error
estimates (see Lovell (\cite{Lov00}) and
http://www.vsop.isas.ac.jp/survey/difwrap).

The $\chi$ errors were computed from the uncertainties in the
$\chi$ calibration for each map added in quadrature to the
rms noise within the $3 \times 3$ pixel box in which $\chi$
was averaged. The $m$ uncertainties were obtained as the
propagated errors from the $I$ and $P$ measurements within
the averaging box, for which the rms noise of $I$ and $P$
was taken as uncertainties of their corresponding measurements.

\section{Discussion}
\label{dis}

\subsection{Kinematics of model components}
\label{kin}

\begin{figure*}
\centering
\includegraphics[bb=0 0 563 363,width=14cm,clip]{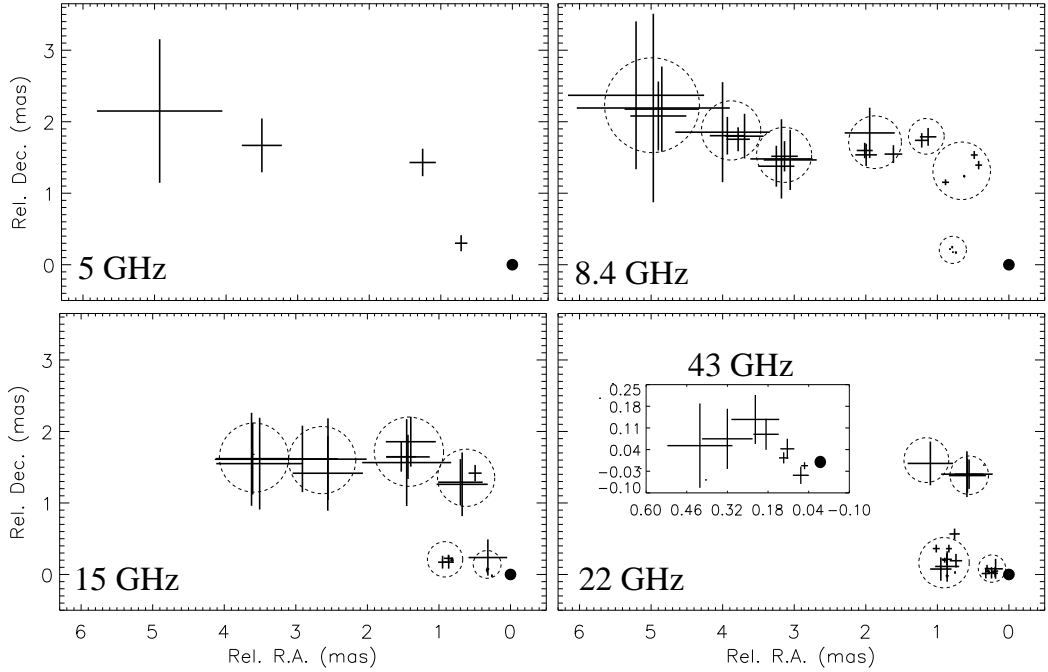}
\caption{Component positions and corresponding uncertainties (symbolized
by the size of the crosses) for the model fits presented in
Tables \ref{fits5} to \ref{fits43}. The dashed circles mark the
various component clumps observed at each frequency. Components with
large uncertainties in the fit parameters (for which no errors are
given in Tables \ref{fits5} to \ref{fits43}) and those with an
uncertain identification (see Tables \ref{fits15} and \ref{fits22})
 are not plotted.}
\label{fitmapall}
\end{figure*}

Figure~\ref{fitmapall} represents the positions of all the model
components in Tables \ref{fits5} to \ref{fits43} within the inner
6\,mas of the jet for which reliable position
estimates were obtained.
This figure shows a clustering of components
around some jet locations, marked by dashed circles.
This clustering was rather stable during a period of four years
and is very unlikely to be produced by chance.
This kind of component distribution was also shown
by G\'omez et al. (\cite{Gom01}) for a different series of data
spanning from 1995 to 1998.
The clustering implies a much slower scenario than reported
for epochs before 1993 (Gabuzda et al.~\cite{Gab94} and references
therein), with proper motions in the range 0.28\,mas\,yr$^{-1}$
($7\,c$) to 0.48\,mas\,yr$^{-1}$ ($12\,c$).
Such proper motions would be observed as systematic shifts of
the positions of the model components by 1.12\,mas to 1.92\,mas
during the time range of our monitoring ($\sim 4$\,yr),
which are not observed in any of the jet regions.

\begin{figure}
\centering
\includegraphics[bb=47 10 662 482,width=9cm,clip]{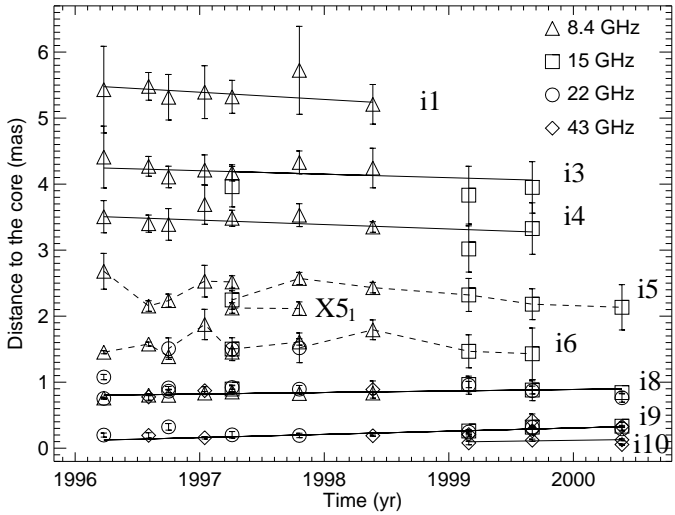}
\caption{Distance to the core as a function of time for
the innermost (within $\aplt 7$\,mas) model components at
8.4\,GHz, 15\,GHz, 22\,GHz and 43\,GHz.
Note that, for the sake of clarity,
only components observed during more than two epochs have been
plotted.  The epochs used are those in Tables
\ref{fits8} to \ref{fits43}, as well as August 1996, January
1997 and May 1998 from G\'omez et al. (2001). Linear
fits of the radial positions for some model components are
also shown (solid lines). The dashed lines connect the
radial positions of components i5, X5$_{1}$ and i6.
Components with large uncertainties in the fit parameters
(for which no errors are given in Tables \ref{fits5} to
\ref{fits43}) and those with an uncertain identification
(see Tables \ref{fits15} and \ref{fits22}) are not plotted.}
\label{fitallrt}
\end{figure}

Figure \ref{fitallrt} presents the distance from the core as a
function of time for the innermost (and most reliable) 8.4\,GHz,
15\,GHz, 22\,GHz and 43\,GHz model components in Tables
\ref{fits5} to \ref{fits43} and those published by
G\'omez et al.~(\cite{Gom01}).
The 5\,GHz data in Table~\ref{fits5} are not plotted due to the
much lower resolution at this frequency, which can produce confusion
in the identification of components.
The evidence of slow proper motions along the jet allowed us to
ensure the most reliable identification of components as the
one which relates a jet feature with those nearest in position
at different observing frequencies and at adjacent observing
epochs.

It is worth to note the significant discrepancy between the
identification of components presented here and the one performed by
Kellermann et al.~(\cite{Kel04}) through their quasi--annual
15\,GHz observations of 0735+178 from 1995 to 2000.
This discrepancy is most probably caused by the poorer time and
frequency sampling of Kellermann et al. and also by the different
model--fitting method used by them, which only considered a maximum
of three jet components per epoch and was performed in the image
plane. Figure~\ref{15ghz} shows that a set of only three jet
components does not suffice to optimally fit the whole VLBI structure
of 0735+178 at 15\,GHz.

Our fits of the radial positions of components
i8, i9 and i10 show slow radial motions outward from the core with
speeds of ($0.023\pm0.004$)\,mas\,yr$^{-1}$,
($0.048\pm0.004$)\,mas\,yr$^{-1}$ and ($0.026\pm0.021$)\,mas\,yr$^{-1}$
([$0.58\pm0.10$]\,$c$, [$1.21\pm0.10$]\,$c$ and [$0.65\pm0.53$]\,$c$),
respectively.
The proper motions of i8 and i9 were interpreted as consistent with
a quasi--stationary behaviour by G\'omez et al. (\cite{Gom01}) for a
shorter time range (from 1995 to 1998) than the one covered by our
observations.
Figure~\ref{fitmapall} shows that these two model components
are more consistently fitted as moving features if our last three
observing epochs (from 1999 to 2000) are taken into account.
This implies that both i8 and i9 slightly speeded up after 1998,
which is more evident for i9 (Fig.~\ref{fitmapall}).
It should also be noted that our estimates provide only the
speeds of the average position of the components during our
observations.
The unaccelerated proper motion of $0.14\pm0.03$\,mas\,yr$^{-1}$
estimated by Homan et al.~(\cite{Hom01}) for i8, through their
bimonthly monitoring program during 1996, accounts for the
irregular motion of this component around its average position.
This rapid, but short distance, motion explains the clumping
of components around their mean positions, which are most likely
produced by the changes of their brightness distributions.

The fits for components i1, i3 and i4 show slight upstream
radial motions. Although these motions could, in principle,
happen in actual relativistic jets under certain conditions
(see e.g. Agudo et al.\ \cite{Agu01}), we prefer to consider
these three components to be quasi-stationary in view of the
large uncertainties in their positions and apparent motions
(with values of [$-0.11\pm0.19$]\,mas\,yr$^{-1}$,
[$-0.05\pm0.09$]\,mas\,yr$^{-1}$ and
[$-0.07\pm0.07$]\,mas\,yr$^{-1}$, respectively), which are in
fact consistent with no motion.
Further evidence for
considering these components as quasi-stationary is based on
their stable structural position angles, with average
values of $66^{\circ}\pm7^{\circ}$, $65^{\circ}\pm5^{\circ}$
and $64^{\circ}\pm5^{\circ}$, respectively, and their stable
total flux densities (see \S~\ref{flu}).

\begin{figure}
\centering
\includegraphics[bb=0 0 685 504,width=9cm,clip]{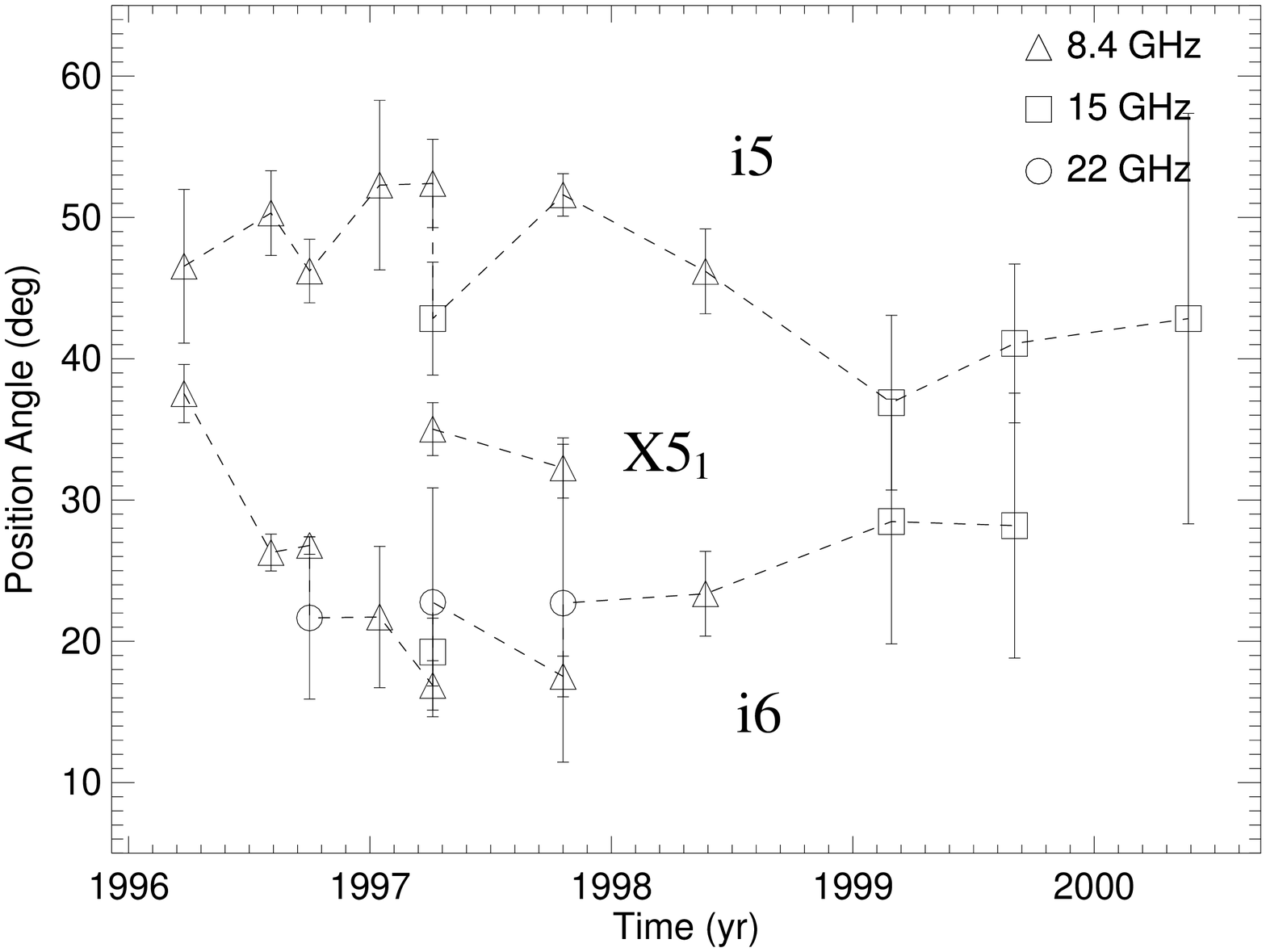}
\caption{Structural position angles as a function
of time for components i5, X5$_{1}$ and i6. The points
shown are the same as those in Fig.~\ref{fitallrt}.}
\label{posang}
\end{figure}

The behaviour of the components related to the second bend, i5 and i6,
is very different. Inspection of Fig.~\ref{15ghz} reveals a
smooth change in the region of the second bend, which
progressively softens from 1997 to 2000.
Figure~\ref{fitmapall} also
shows a significant dispersion for the position of component
i6 (from March 1996 to October 1997), which is larger than its
typical position uncertainty (especially at 8.4\,GHz).
Fig. \ref{posang} represents the structural position angles of
components i5, X5$_{1}$ and i6 during the time covered by our
observations. We can see that i6 moved from a structural position
angle of $\sim 38^{\circ}$ at the beginning of 1996 to
$\sim 20^{\circ}$ from the end of 1996 to the beginning of 1997.
During 1999, it returned to $\sim 30^{\circ}$, then finally
became undetectable in our last observing epoch (May 2000).
Although the variations are less significant, the structural position 
angle of i5 also changed from $\sim 47^{\circ}$ at the beginning
of 1996, to $\sim 52^{\circ}$, during 1997, to
$\sim 37^{\circ}$ at the beginning of 1999, and then back to
almost the initial position angle at the beginning of 2000.
During the time period of maximum angular separation between i5
and i6 (during 1997) a new component (X5$_{1}$) was fitted at
intermediate angular positions of $35^{\circ} \pm 2^{\circ}$
(in April 1997) and $32^{\circ} \pm 2^{\circ}$ (in October 1997).
Although it was necessary to introduce component X5$_{1}$ for
these two epochs in order to obtain a good model fit, it
seems very unlikely that this is related to the appearance of a
new real feature in the jet. We consider X5$_{1}$ to represent
emission from the underlying jet in the gap region produced as
a consequence of the angular separation of i5 and i6.

The behaviour of i5 and i6 between 1996 and 2000 is related to
a ``softening'' of the second bend, which is consistent with the
fairly rectilinear east--north--east jet structure reported
by the 15\,GHz image of Lister \& Homan~(\cite{Lis05}) taken
on November 2002.
This softening is produced by an apparent backward motion
of component i6.
Again, this should not be understood as a
physical motion of jet plasma back to the core, but rather as
induced by a decrease in the total flux density of the bent area,
shifting the peak of the emission (associated with the component)
to positions closer to the core until component i6 became
undetectable by May 2000 (see Tables \ref{fits5} to \ref{fits43}
and \S~\ref{flu}).

\begin{table}
\caption[]{5\,GHz model fits for 6 April 1997.}
\begin{flushleft}
\scriptsize
\begin{tabular} {lcccccc}
\hline\noalign{\smallskip}
Comp. &   $I$ &  $r$ &    $\theta$ &  $FWHM$ & $\chi$ & $m$\\
      & (mJy) & (mas) & ($^{\circ}$) & (mas)  & ($^{\circ}$) & $\%$  \\
\hline\noalign{\smallskip}
Core  & 249$\pm$12 &  0.00           & 0         &       $<$0.43  & -23$\pm$3  &  2.2$\pm$0.3\\
C8    & 455$\pm$8  &  0.78$\pm$0.09  & 67$\pm$30 & 0.82$\pm$0.08  & -14$\pm$3  &  4.1$\pm$0.4\\
C5+6  & 100$\pm$13 &  1.90$\pm$0.15  & 41$\pm$5  & 1.55$\pm$0.23  &  -1$\pm$3  &   10$\pm$2  \\
C3+4  &  47$\pm$11 &  3.88$\pm$0.15  & 65$\pm$6  & 1.18$\pm$0.51  & -11$\pm$4  &    8$\pm$5  \\
C1    &  19$\pm$9  &  5.38$\pm$0.60  & 66$\pm$10 & 1.59$\pm$0.93  &    ...     &       ...   \\
CE1   &  70$\pm$40 & 10.33$\pm$1.19  & 73$\pm$6  & 6.06$\pm$1.67  &    ...     &       ...   \\
CE2   &   $\sim$81 &     $\sim$21.79 &  $\sim$76 &    $\sim$16.40 &    ...     &       ...   \\
\noalign{\smallskip}
\hline
\end{tabular}
\end{flushleft}
\label{fits5}
\end{table}

\begin{table}
\caption[]{8.4\,GHz model fits.}
\begin{flushleft}
\scriptsize
\begin{tabular} {lcccccc}
\hline\noalign{\smallskip}
Comp. &   $I$ &  $r$ &    $\theta$ &  $FWHM$ & $\chi$ & $m$\\
      & (mJy) & (mas) & ($^{\circ}$) & (mas)  & ($^{\circ}$) & $\%$  \\
\hline\noalign{\smallskip}
\multicolumn{7}{c}{26 March 1996}\\
\hline\noalign{\smallskip}
Core & 590$\pm$39 &  0.00          & 0         & 0.20$\pm$0.11  &    ...     &    ...   \\
X8   & 629$\pm$24 &  0.76$\pm$0.01 & 77$\pm$2  & 0.42$\pm$0.06  &    ...     &    ...   \\
X6   & 161$\pm$24 &  1.45$\pm$0.02 & 38$\pm$2  & 0.91$\pm$0.22  &-0.8$\pm$3  & 7$\pm$1  \\
X5   &  16$\pm$9  &  2.68$\pm$0.27 & 47$\pm$5  &       $<$0.54  &   5$\pm$3  & 23$\pm$11\\
X4   &  23$\pm$12 &  3.51$\pm$0.24 & 65$\pm$9  &       $<$0.61  & -16$\pm$3  & 14$\pm$8 \\
X3   &  21$\pm$15 &  4.41$\pm$0.47 & 65$\pm$8  &       $<$1.21  &    ...     &    ...   \\
X1   &  20$\pm$15 &  5.43$\pm$0.66 & 66$\pm$13 &       $<$2.10  &    ...     &    ...   \\
XE   &   $\sim$77 &     $\sim$11.08&  $\sim$72 &    $\sim$8.06  &    ...     &    ...   \\

\hline\noalign{\smallskip}
\multicolumn{7}{c}{2 October 1996}\\
\hline\noalign{\smallskip}
Core & 396$\pm$4  &  0.00          & 0         & 0.25$\pm$0.04 & -47$\pm$3  &  2.4$\pm$0.2\\
X8   & 434$\pm$6  &  0.80$\pm$0.01 & 77$\pm$1  & 0.53$\pm$0.02 & -49$\pm$3  &  2.7$\pm$0.3\\
X6   & 110$\pm$11 &  1.39$\pm$0.02 & 27$\pm$1  & 0.86$\pm$0.03 & -15$\pm$3  &  8.6$\pm$0.8\\
X5   &  36$\pm$8  &  2.24$\pm$0.10 & 46$\pm$2  & 0.99$\pm$0.15 &  13$\pm$3  &  25$\pm$5   \\
X4   &  15$\pm$8  &  3.39$\pm$0.24 & 64$\pm$7  & 1.00$\pm$0.43 & -21$\pm$3  &  19$\pm$6   \\
X3   &  18$\pm$8  &  4.11$\pm$0.16 & 64$\pm$4  & 0.72$\pm$0.43 & -36$\pm$3  &  17$\pm$6   \\
X1   &  19$\pm$12 &  5.32$\pm$0.35 & 66$\pm$6  & 1.26$\pm$0.53 &     ...    &    ...      \\
XE   &   $\sim$62 &     $\sim$10.8 &  $\sim$73 &    $\sim$7.09 &     ...    &    ...      \\

\hline\noalign{\smallskip}
\multicolumn{7}{c}{6 April 1997}\\
\hline\noalign{\smallskip}
Core     & 350$\pm$4  &  0.00          & 0        & 0.22$\pm$0.03 &  -3$\pm$4 &  1.1$\pm$0.3\\
X8       & 310$\pm$6  &  0.85$\pm$0.01 & 75$\pm$1 & 0.67$\pm$0.02 & -25$\pm$3 &  3.9$\pm$0.6\\
X6       &  52$\pm$8  &  1.46$\pm$0.05 & 17$\pm$2 & 0.60$\pm$0.10 & -11$\pm$3 & 15.8$\pm$1.7\\
X5$_{1}$ &  31$\pm$7  &  2.12$\pm$0.08 & 35$\pm$2 & 0.67$\pm$0.15 &   1$\pm$3 &   21$\pm$4  \\
X5$_{2}$ &  19$\pm$5  &  2.52$\pm$0.09 & 52$\pm$3 & 0.53$\pm$0.34 &   2$\pm$3 &   19$\pm$6  \\
X4       &  23$\pm$7  &  3.48$\pm$0.12 & 64$\pm$3 & 1.08$\pm$0.26 & -11$\pm$3 &   17$\pm$7  \\
X3       &  12$\pm$4  &  4.17$\pm$0.12 & 65$\pm$2 &       $<$0.56 & -23$\pm$3 &   14$\pm$8  \\
X1       &  17$\pm$8  &  5.32$\pm$0.25 & 67$\pm$5 & 1.39$\pm$0.73 &     ...   &      ...    \\
XE       &  $\sim$56  &     $\sim$11.04&  $\sim$73 &   $\sim$6.41 &     ...   &      ...    \\

\hline\noalign{\smallskip}
\multicolumn{7}{c}{18 October 1997}\\
\hline\noalign{\smallskip}
Core     & 391$\pm$2  &  0.00          & 0         & 0.18$\pm$0.01 &  80$\pm$3 &  3.2$\pm$0.2\\
X8       & 222$\pm$2  &  0.83$\pm$0.01 & 73$\pm$1  & 0.71$\pm$0.03 & -84$\pm$3 &  4.9$\pm$0.5\\
X6       &  40$\pm$4  &  1.61$\pm$0.05 & 18$\pm$1  & 0.51$\pm$0.18 & -52$\pm$3 & 14.5$\pm$1.4\\
X5$_{1}$ &  28$\pm$4  &  2.11$\pm$0.10 & 32$\pm$2  & 0.48$\pm$0.23 & -34$\pm$3 & 24.3$\pm$2.2\\
X5$_{2}$ &  37$\pm$4  &  2.57$\pm$0.09 & 52$\pm$2  & 0.78$\pm$0.24 & -20$\pm$3 &   23$\pm$3  \\
X4       &  18$\pm$7  &  3.53$\pm$0.17 & 67$\pm$4  & 0.76$\pm$0.44 & -44$\pm$3 &   14$\pm$5  \\
X3       &  18$\pm$6  &  4.33$\pm$0.17 & 65$\pm$3  & 0.75$\pm$0.44 & -47$\pm$3 &   16$\pm$5  \\
X1       &  10$\pm$9  &  5.72$\pm$0.67 & 66$\pm$9  & 1.37$\pm$0.64 &     ...   &      ...    \\
XE       &  $\sim$67  &     $\sim$11.28&  $\sim$73 &    $\sim$6.98 &     ...  &      ...     \\

\noalign{\smallskip}
\hline
\end{tabular}
\end{flushleft}
\label{fits8}
\end{table}

\begin{table}
\caption[]{15\,GHz model fits.}
\begin{flushleft}
\scriptsize
\begin{tabular} {lcccccc}
\hline\noalign{\smallskip}
Comp. &   $I$ &  $r$ &    $\theta$ &  $FWHM$ & $\chi$ & $m$\\
      & (mJy) & (mas) & ($^{\circ}$) & (mas)  & ($^{\circ}$) & $\%$  \\
\hline\noalign{\smallskip}
\multicolumn{7}{c}{6 April 1997}\\
\hline\noalign{\smallskip}
Core   &   292$\pm$7  &  0.00          & 0         &  0.15$\pm$0.02 &  -9$\pm$4 &  0.6$\pm$0.5\\
U9$^1$ &    15$\pm$8  &  0.40$\pm$0.23 & 53$\pm$23 &        $<$0.38 &     ...   &      ...    \\
U8     &   236$\pm$18 &  0.90$\pm$0.06 & 75$\pm$2  &  0.67$\pm$0.03 & -39$\pm$3 &  2.6$\pm$1.3\\
U6     &    47$\pm$9  &  1.50$\pm$0.10 & 19$\pm$2  &  0.69$\pm$0.18 & -25$\pm$3 &   18$\pm$7  \\
U5     &    24$\pm$12 &  2.25$\pm$0.14 & 43$\pm$4  &  0.76$\pm$0.34 &  -1$\pm$3 &   22$\pm$15 \\
U4$^1$ &    10$\pm$10 &  2.92$\pm$0.33 & 61$\pm$8  &  0.53$\pm$0.39 &     ...   &      ...    \\
U3     &    11$\pm$9  &  3.96$\pm$0.31 & 66$\pm$8  &      $\pm$1.15 &     ...   &      ...    \\
UE     &     $\sim$14 &     $\sim$4.83 & $\sim$64  &     $\sim$1.45 &     ...   &      ...    \\

\hline\noalign{\smallskip}
\multicolumn{7}{c}{27 February 1999}\\
\hline\noalign{\smallskip}

Core &  417$\pm$2  &  0.00          & 0         &  0.09$\pm$0.02 & -69$\pm$3 &  1.3$\pm$0.2\\
U9   &   83$\pm$4  &  0.26$\pm$0.01 & 94$\pm$5  &  0.22$\pm$0.13 & -68$\pm$3 &  2.0$\pm$0.4\\
U8   &   47$\pm$9  &  0.96$\pm$0.05 & 80$\pm$5  &  0.55$\pm$0.18 & -46$\pm$3 &   11$\pm$4  \\
U6   &   25$\pm$15 &  1.47$\pm$0.25 & 29$\pm$9  &  0.91$\pm$0.32 &     ...   &       ...  \\
U5   &   13$\pm$9  &  2.32$\pm$0.25 & 37$\pm$6  &  0.60$\pm$0.44 & -11$\pm$3 &   36$\pm$17\\
U4   &   18$\pm$13 &  3.02$\pm$0.35 & 58$\pm$9  &  1.08$\pm$0.66 &     ...   &       ...  \\
U3   &   13$\pm$13 &  3.83$\pm$0.44 & 66$\pm$8  &  1.03$\pm$1.01 &     ...   &       ...  \\
UE   &    $\sim$6  &     $\sim$5.31 &  $\sim$65 &     $\sim$1.26 &     ...   &       ...  \\

\hline\noalign{\smallskip}
\multicolumn{7}{c}{1 September 1999}\\
\hline\noalign{\smallskip}
Core &   365$\pm$4  &  0.00          & 0         & 0.12$\pm$0.03 & -38$\pm$5 &  3.6$\pm$0.3\\
U9   &   176$\pm$5  &  0.32$\pm$0.01 & 76$\pm$3  & 0.24$\pm$0.07 & -39$\pm$5 &  7.0$\pm$0.7\\
U8   &    38$\pm$8  &  0.88$\pm$0.06 & 79$\pm$6  & 0.57$\pm$0.16 & -44$\pm$5 &   13$\pm$6  \\
U6   &     9$\pm$8  &  1.43$\pm$0.39 & 28$\pm$9  & 0.48$\pm$0.44 &     ...   &      ...    \\
U5   &    24$\pm$9  &  2.18$\pm$0.23 & 41$\pm$6  & 0.89$\pm$0.28 &  -4$\pm$5 &   21$\pm$18 \\
U4   &    17$\pm$10 &  3.33$\pm$0.39 & 61$\pm$6  & 0.92$\pm$0.44 &     ...   &      ...    \\
U3   &     8$\pm$6  &  3.95$\pm$0.39 & 66$\pm$6  &       $<$1.31 &     ...   &      ...    \\
UE   &     $\sim$6  &     $\sim$5.04 &  $\sim$67 &    $\sim$1.20 &     ...   &      ...    \\

\hline\noalign{\smallskip}
\multicolumn{7}{c}{20 May 2000}\\
\hline\noalign{\smallskip}
Core &   523$\pm$6  &  0.00          & 0         & 0.12$\pm$0.02 & -60$\pm$5 &   2.0$\pm$0.2\\
U9   &   144$\pm$9  &  0.33$\pm$0.01 & 79$\pm$7  & 0.30$\pm$0.05 & -63$\pm$5 &   2.9$\pm$0.6\\
U8   &    86$\pm$8  &  0.84$\pm$0.02 & 76$\pm$3  & 0.54$\pm$0.05 & -70$\pm$5 &    10$\pm$3  \\
U5   &    30$\pm$17 &  2.13$\pm$0.34 & 43$\pm$15 & 1.31$\pm$0.91 &    ...    &      ...     \\
U4   &     $\sim$7  &     $\sim$3.16 &  $\sim$62 &    $\sim$0.57 &    ...    &      ...     \\
U3   &     $\sim$15 &     $\sim$3.97 &  $\sim$66 &    $\sim$0.88 &    ...    &      ...     \\
UE   &     $\sim$3  &     $\sim$5.72 &  $\sim$69 &    $\sim$1.03 &    ...    &      ...     \\

\noalign{\smallskip}
\hline
\end{tabular}
\end{flushleft}
$^1$Uncertain identification of this model component.
\label{fits15}
\end{table}

\begin{table}
\caption[]{22\,GHz model fits.}
\begin{flushleft}
\scriptsize
\begin{tabular} {lcccccc}
\hline\noalign{\smallskip}
Comp. &   $I$ &  $r$ &    $\theta$ &  $FWHM$ & $\chi$ & $m$\\
      & (mJy) & (mas) & ($^{\circ}$) & (mas)  & ($^{\circ}$) & $\%$  \\
\hline\noalign{\smallskip}
\multicolumn{7}{c}{26 March 1996}\\
\hline\noalign{\smallskip}

Core     &   353$\pm$9   & 0.00          & 0         &        $<$0.10 &     ...     &    ...  \\
K9       &    73$\pm$9   & 0.20$\pm$0.03 & 76$\pm$18 &        $<$0.32 &     ...     &    ...  \\
K8$_{1}$ &   232$\pm$13  & 0.75$\pm$0.01 & 88$\pm$1  &  0.34$\pm$0.04 &  60$\pm$4   &  4$\pm$3\\
K7       &   122$\pm$16  & 0.95$\pm$0.05 & 53$\pm$4  &  0.68$\pm$0.10 &     ...     &    ...  \\
K8$_{2}$ &    93$\pm$13  & 1.08$\pm$0.04 & 70$\pm$3  &  0.34$\pm$0.07 &     ...     &    ...  \\
K6       &     $\sim$5   &    $\sim$1.57 &  $\sim$13 &     $\sim$0.01 &     ...     &    ...  \\
K5$^1$   &    45$\pm$34  & 1.90$\pm$0.20 & 35$\pm$8  &  1.08$\pm$0.34 &     ...     &    ...  \\
K4       &     $\sim$8   &    $\sim$3.12 &  $\sim$61 &     $\sim$0.68 &     ...     &    ...  \\
K3       &     $\sim$10  &    $\sim$4.07 &  $\sim$65 &     $\sim$0.67 &     ...     &    ...  \\
K1       &     $\sim$12  &    $\sim$5.13 &  $\sim$64 &     $\sim$0.90 &     ...     &    ...  \\

\hline\noalign{\smallskip}
\multicolumn{7}{c}{2 October 1996}\\
\hline\noalign{\smallskip}

Core     &  217$\pm$7  & 0.00          & 0         &      $<$0.10   &  14$\pm$3 &  2.2$\pm$1.2\\
K9       &   33$\pm$8  & 0.32$\pm$0.05 & 88$\pm$12 &  0.26$\pm$0.05 &     ...   &      ...    \\
K8$_{1}$ &   70$\pm$10 & 0.86$\pm$0.02 & 91$\pm$2  &  0.30$\pm$0.09 & -72$\pm$3 &    6$\pm$4  \\
K8$_{2}$ &  153$\pm$10 & 0.91$\pm$0.02 & 67$\pm$3  &  0.59$\pm$0.08 &     ...   &      ...    \\
K6       &   26$\pm$14 & 1.51$\pm$0.16 & 22$\pm$6  &  0.68$\pm$0.31 &  -9$\pm$3 &   34$\pm$30 \\
K5       &    $\sim$22 &    $\sim$2.31 &  $\sim$47 &     $\sim$1.28 &     ...   &      ...    \\

\hline\noalign{\smallskip}
\multicolumn{7}{c}{6 April 1997}\\
\hline\noalign{\smallskip}

Core     &  184$\pm$11 & 0.00          & 0         &        $<$0.18 &  -6$\pm$4 &  2.7$\pm$1.4\\
K9       &   47$\pm$11 & 0.20$\pm$0.05 & 67$\pm$45 &        $<$0.35 &   5$\pm$3 &  4.9$\pm$2.4\\
K8       &  171$\pm$28 & 0.92$\pm$0.03 & 77$\pm$2  &  0.69$\pm$0.06 &      ...  &       ...   \\
K6       &   43$\pm$24 & 1.50$\pm$0.17 & 23$\pm$8  &  0.79$\pm$0.20 &   4$\pm$3 &   36$\pm$29 \\
K5       &    $\sim$19 &    $\sim$2.41 & $\sim$46  &     $\sim$0.93 &  $\sim$-2 &   $\sim$86 \\
KE$_{1}$ &    $\sim$18 &    $\sim$4.08 & $\sim$64  &     $\sim$1.09 &     ...   &       ...   \\

\hline\noalign{\smallskip}
\multicolumn{7}{c}{18 October 1997}\\
\hline\noalign{\smallskip}

Core     &  244$\pm$6  & 0.00          & 0         &        $<$0.13 &  18$\pm$3 &  5.2$\pm$0.5\\
K9       &   43$\pm$8  & 0.19$\pm$0.03 & 85$\pm$17 &        $<$0.21 &  38$\pm$4 &  6.0$\pm$1.1\\
K8       &  107$\pm$16 & 0.89$\pm$0.06 & 76$\pm$5  &  0.69$\pm$0.05 &     ...   &      ...    \\
K6       &   34$\pm$20 & 1.52$\pm$0.23 & 23$\pm$11 &  0.79$\pm$0.35 &     ...   &      ...    \\
K5       &    $\sim$25 &    $\sim$2.36 &  $\sim$42 &     $\sim$1.08 &     ...   &      ...    \\
KE$_{1}$ &    $\sim$20 &    $\sim$3.87 &  $\sim$65 &     $\sim$1.37 &     ...   &      ...    \\

\hline\noalign{\smallskip}
\multicolumn{7}{c}{27 February 1999}\\
\hline\noalign{\smallskip}

Core     &  414$\pm$9  & 0.00          & 0         &        $<$0.19 & -42$\pm$3 & 0.8$\pm$0.4\\
K9       &   76$\pm$18 & 0.24$\pm$0.03 & 85$\pm$23 &  0.25$\pm$0.11 &     ...   &     ...    \\
K8       &   33$\pm$28 & 0.95$\pm$0.14 & 85$\pm$10 &  0.49$\pm$0.24 &     ...   &     ...    \\
KE$_{2}$ &    $\sim$60 &    $\sim$1.80 &  $\sim$42 &     $\sim$1.73 &     ...   &     ...    \\

\hline\noalign{\smallskip}
\multicolumn{7}{c}{1 September 1999}\\
\hline\noalign{\smallskip}

Core     &  345$\pm$11 & 0.00          & 0         &  0.09$\pm$0.02 & -32$\pm$3 & 2.1$\pm$0.6\\
K9       &  145$\pm$14 & 0.32$\pm$0.02 & 75$\pm$11 &  0.26$\pm$0.10 & -29$\pm$3 & 7.3$\pm$1.7\\
K8       &   24$\pm$18 & 0.87$\pm$0.15 & 83$\pm$14 &  0.51$\pm$0.26 &     ...   &     ...    \\
KE$_{2}$ &    $\sim$47 &    $\sim$2.31 &  $\sim$49 &     $\sim$2.06 &     ...   &     ...    \\

\hline\noalign{\smallskip}
\multicolumn{7}{c}{20 May 2000}\\
\hline\noalign{\smallskip}

Core     &  520$\pm$7  & 0.00          & 0         & 0.12$\pm$0.01 & -52$\pm$3 &  1.7$\pm$0.3\\
K9       &  134$\pm$19 & 0.29$\pm$0.02 & 81$\pm$5  & 0.31$\pm$0.07 & -78$\pm$3 &  3.3$\pm$1.6\\
K8       &   92$\pm$16 & 0.76$\pm$0.07 & 75$\pm$6  & 0.61$\pm$0.09 &     ...   &      ...    \\
KE$_{2}$ &    $\sim$42 &    $\sim$2.91 &  $\sim$57 &    $\sim$2.27 &     ...   &      ...    \\

\noalign{\smallskip}
\hline
\end{tabular}
\end{flushleft}
$^1$Uncertain identification of this model component.
\label{fits22}
\end{table}

\begin{table}
\caption[]{43\,GHz model fits.}
\begin{flushleft}
\scriptsize
\begin{tabular} {lcccccc}
\hline\noalign{\smallskip}
Comp. &   $I$ &  $r$ &    $\theta$ &  $FWHM$ & $\chi$ & $m$\\
      & (mJy) & (mas) & ($^{\circ}$) & (mas)  & ($^{\circ}$) & $\%$  \\

\hline\noalign{\smallskip}
\multicolumn{7}{c}{27 February 1999}\\
\hline\noalign{\smallskip}

Core     &   315$\pm$12  &  0.00           & 0          &  0.07$\pm$0.01 & ... & ... \\
Q10      &    68$\pm$9   &  0.08$\pm$0.03  & 123$\pm$13 &        $<$0.22 & ... & ... \\
Q9$_{1}$ &    62$\pm$27  &  0.21$\pm$0.02  &  64$\pm$14 &  0.21$\pm$0.04 & ... & ... \\
Q9$_{2}$ &     $\sim$7   &     $\sim$0.40  &   $\sim$98 &     $\sim$0.02 & ... & ... \\
Q8$_{1}$ &     $\sim$5   &     $\sim$0.62  &   $\sim$92 &     $\sim$0.10 & ... & ... \\
Q8$_{2}$ &     $\sim$11  &     $\sim$0.79  &   $\sim$75 &     $\sim$0.28 & ... & ... \\
Q8$_{3}$ &     $\sim$8   &     $\sim$1.08  &   $\sim$86 &     $\sim$0.22 & ... & ... \\

\hline\noalign{\smallskip}
\multicolumn{7}{c}{1 September 1999}\\
\hline\noalign{\smallskip}

Core      &  257$\pm$14 & 0.00          & 0         &         $<$0.08 & ... & ... \\
Q10       &   79$\pm$18 & 0.12$\pm$0.02 & 70$\pm$21 &         $<$0.20 & ... & ... \\
Q9$_{1}$  &   43$\pm$29 & 0.26$\pm$0.06 & 58$\pm$14 &         $<$0.34 & ... & ... \\
Q9$_{2}$  &   59$\pm$43 & 0.42$\pm$0.10 & 83$\pm$19 &   0.29$\pm$0.22 & ... & ... \\

\hline\noalign{\smallskip}
\multicolumn{7}{c}{20 May 2000}\\
\hline\noalign{\smallskip}

Core     &  252$\pm$12 &  0.00           & 0          &  0.07$\pm$0.04 & -51$\pm$7 &  1.4$\pm$0.5\\
Q11      &  163$\pm$9  &  0.06$\pm$0.01  & 103$\pm$25 &        $<$0.08 & -59$\pm$8 &  1.4$\pm$0.6\\
Q10      &   82$\pm$12 &  0.13$\pm$0.01  &  84$\pm$11 &        $<$0.21 & -74$\pm$8 &  3.1$\pm$1.1\\
Q9       &  162$\pm$34 &  0.33$\pm$0.08  &  77$\pm$16 &  0.47$\pm$0.07 &     ...   &       ...   \\
Q8$_{1}$ &    $\sim$14 &     $\sim$0.76  &   $\sim$71 &     $\sim$0.20 &     ...   &       ...   \\
Q8$_{2}$ &    $\sim$18 &     $\sim$1.10  &   $\sim$81 &     $\sim$0.32 &     ...   &       ...   \\

\noalign{\smallskip}
\hline
\end{tabular}
\end{flushleft}
\label{fits43}
\end{table}

\subsection{Flux density evolution of model components}
\label{flu}

\begin{figure}
\centering
\includegraphics[bb=0 0 535 573,width=9cm,clip]{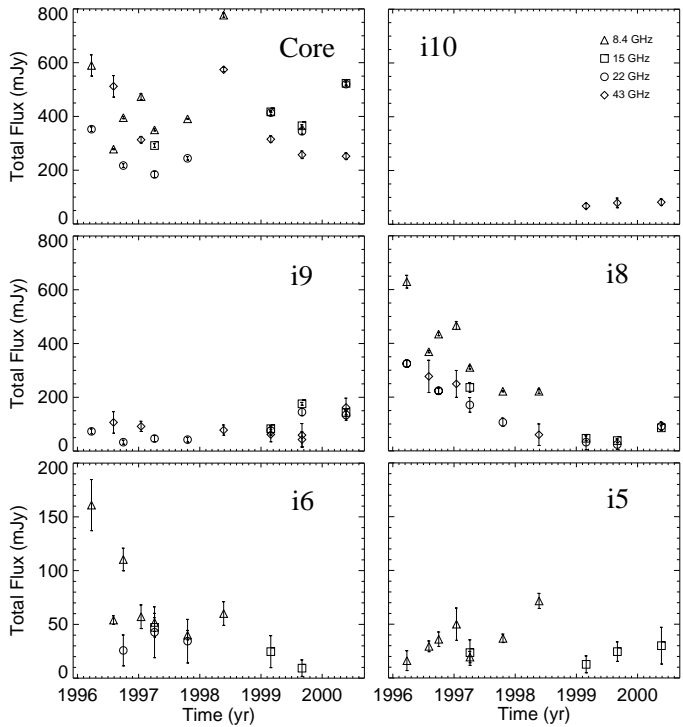}
\caption{5\,GHz, 8.4\,GHz, 15\,GHz and 22\,GHz and 43\,GHz
total flux density evolution for the Core, i10, i9, i8, i6 and
i5 components.
The epochs shown
are those in Table \ref{datlst}, as well as August 1996, January
1997 and May 1998 from G\'omez et al. (2001). Each frequency
is represented by a different symbol, as indicated in the right
top inset. Note the different total flux density scale used for the
plots of i6 and i5.}
\label{fluxev}
\end{figure}

Fig. \ref{fluxev} presents the 5\,GHz, 8.4\,GHz, 15\,GHz and 22\,GHz
total flux density evolution for the core and components i10, i9, i8,
i6 and i5 during the time span covered by our observations.
Components not included in this figure, X5$_{1}$; i4; i3 and i1,
showed little 8.4\,GHz total flux density evolution with weighted means
of ($29\pm4$)\,mJy, ($17\pm4$)\,mJy ([$17\pm8$]\,mJy at 15\,GHz),
($16\pm2$)\,mJy ([$9\pm5$]\,mJy at 15\,GHz) and ($11\pm3$)\,mJy,
respectively.

This figure shows that the total flux density of i6 at 8.4\,GHz
decreased from $\sim 160$\,mJy at the beginning of 1996 to $\sim
10$\,mJy at the end of 1999, while the nearby
feature i5 displayed a quite stable total flux density of
($35\pm6$)\,mJy at 8.4\,GHz and ($20\pm5$)\,mJy at 15\,GHz;
i5 was still detected in May 2000, whereas i6 had become
undetectable by that time.

A similar trend as for i6 was observed for component
i8 at the first bend, which quasi-monotonically decreased its
total flux density by $\sim 500$\,mJy at 8.4\,GHz during the same
period.
Simultaneously, the total flux density of the nearby component i9
remained almost constant at $\sim 100$ mJy from 1996 to 1999.
Component i10 likewise maintained a very stable total flux density
of ($74\pm7$)\,mJy at 43\,GHz from February 1999 to May 2000.

While the total flux density evolution of i8, i9 and i10 was
monotonic for most of the time between 1996 and 2000, the core
displayed significant rapid fluctuations on typical time--scales
of several weeks and with amplitudes of $\sim 20$\,\% around
$400$\,mJy.

\begin{figure}
\centering
\includegraphics[bb=0 0 645 468,width=8cm,clip]{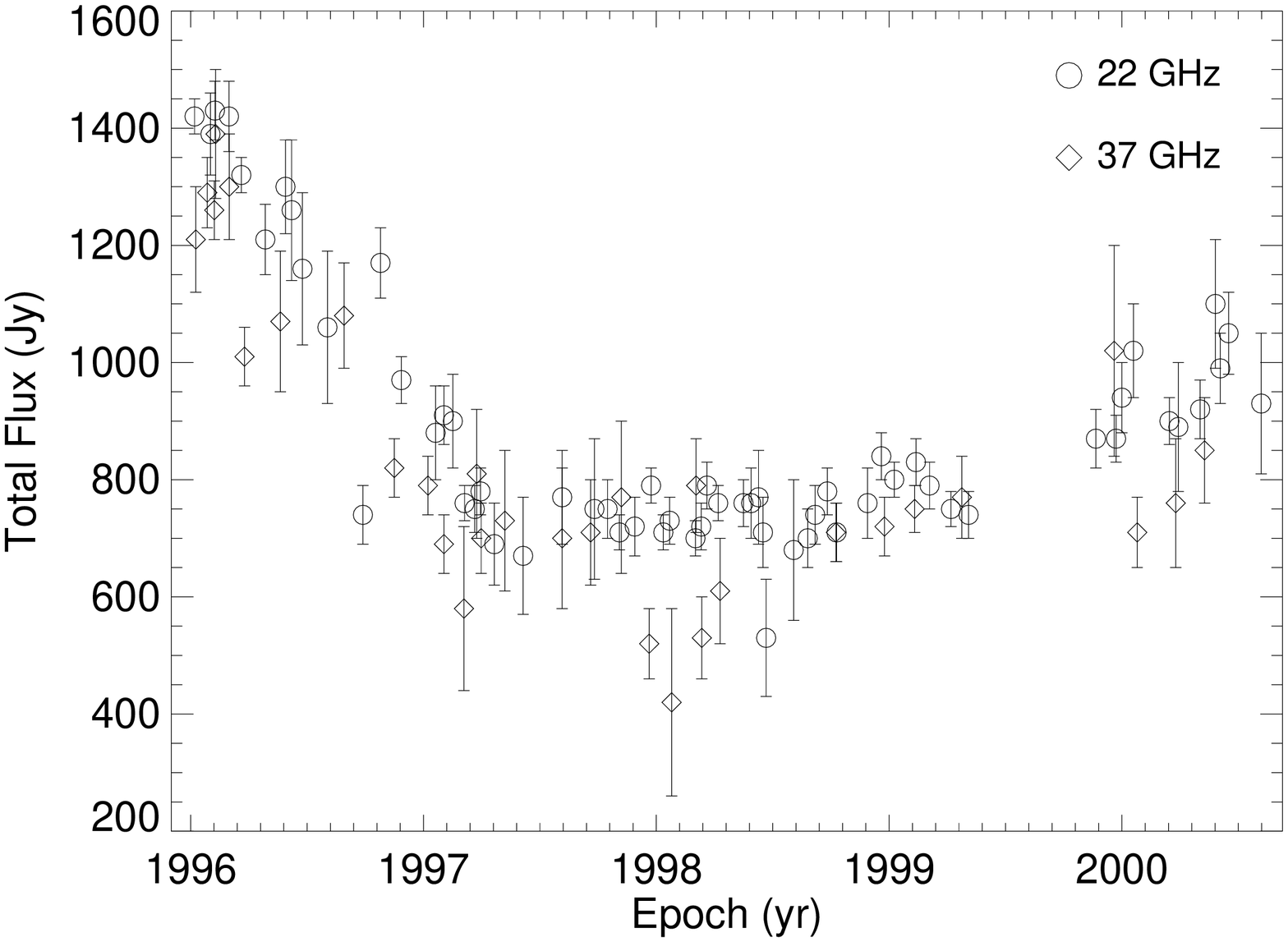}
\caption{Integrated 22\,GHz and 37\,GHz total flux density light curves
of 0735+178 from 1996 to 2000.
Data obtained by the monitoring program of extragalactic radio sources of
the Mets\"ahovi radio observatory (Ter\"asranta et al.~\cite{Ter04}).}
\label{lcurv}
\end{figure}

Note that, whereas the total flux density of the core fluctuated
and the flux densities of i9 and i10 remained almost constant,
the behaviour of i8 coincided with the single--dish integrated
total flux density evolution of 0735+178 between early 1996 and
late 1997 at 4.8\,GHz, 8\,GHz and 14.5\,GHz (Aller et al.\ \cite{All99})
and at 22\,GHz and 37\,GHz (Fig.\,\ref{lcurv} and
Ter\"asranta et al. \cite{Ter04}).
The data from both single dish monitoring programs showed a monotonic
decrease in total flux density of $\sim 500$\,mJy at all frequencies
during this time range, which is precisely the amount of flux density
decrease observed for i8.

This, together with the fact that i8, i9, i10 and the
core are typically responsible for more than the 90\,\% of the total
emission of the source at high frequencies (22\,GHz and 43\,GHz)
and more than 80\,\% at lower frequencies, shows that the decrease
in total flux density measured by the single--dish monitoring was
determined primarily by the flux density decrease of i8.

\subsection{Doppler factor and geometry of the inner jet}
\label{dop}

The highly bent VLBI structure of 0735+178 suggests that the
source is oriented at a small angle to our line
of sight. This is also supported by the large superluminal
motions measured in the jet before 1993, and by the observed high
degree of variability for the core (see \S \ref{flu}).

It is possible to constrain this
angle using our proper motion measurements and previous
estimates of the jet's Doppler factor\footnote{
$\delta=[\Gamma(1-\beta cos \theta)]^{-1}$ where $\Gamma=(1-\beta^2)
^{-1/2}$ is the Lorentz factor, $\beta$ is the speed in units of
the speed of light and $\theta$ is the angle between the
direction of the flow and the line of sight.} ($\delta$).
G\"uijosa \& Daly (\cite{Gui96}) computed a $\delta_{eq} \approx 7$
(estimation of the Doppler factor under the assumption that the
magnetic field and the relativistic particles of the plasma are in
equipartition). Using different assumptions, L\"ahteenm\"aki \&
Valtaoja (\cite{Lah99}) reported $\delta_{var} \approx 4$ (Doppler
factor estimated from the variability of the long term radio light
curves).
Other estimates
of $\delta$ for 0735+178 in the literature lie between the previous
values of $\delta_{eq}$ and $\delta_{var}$.
To test these values, we have computed the variability
Doppler factor $\delta_{var}$ of the core in 0735+178 through the expression
defined by Jorstad et al. (\cite{Jor05}) and using our own data. The result
gives $\delta_{var} \approx 5$, which agrees well with the
previous estimates.

We then assume that the true $\delta$ for the innermost
jet in 0735+178 ranges between 4 and 7, and use our
observed proper motions of components i8, i9 and i10 (which
essentially range between 0.1$\,c$ and 1.3$\,c$) to
estimate the possible ranges of viewing angles
($\theta$) and plasma bulk Lorentz factors ($\Gamma$) for the first
mas of the jet, using the relation
$\beta_{ap}=(\beta sin \theta)/(1-\beta cos \theta)$ where
$\beta_{ap}$ is the measured apparent proper motion (e.g., Pearson
\& Zensus \cite{Pea87}) together with the definition of the Doppler factor
(see above). This constrains $\theta$ to values between $0^\circ$ and
$9^\circ$, as small as expected, and constrains $\Gamma$ to values
between $2$ and $4$.
Note that the speeds used for these estimates are associated
with jet features that are related to bends (at least for i8;
see Figs. \ref{22ghz} and \ref{43ghz}) and we have not observed
fast superluminal speeds such as those reported before 1993.
Therefore, our estimates of $\theta$ and $\Gamma$ should be related
to the pattern speeds of the innermost bent jet structure.

\subsection{Polarization evolution of model components}
\label{pol}

\begin{figure}
\centering
\includegraphics[bb=0 0 447 535,width=9cm,clip]{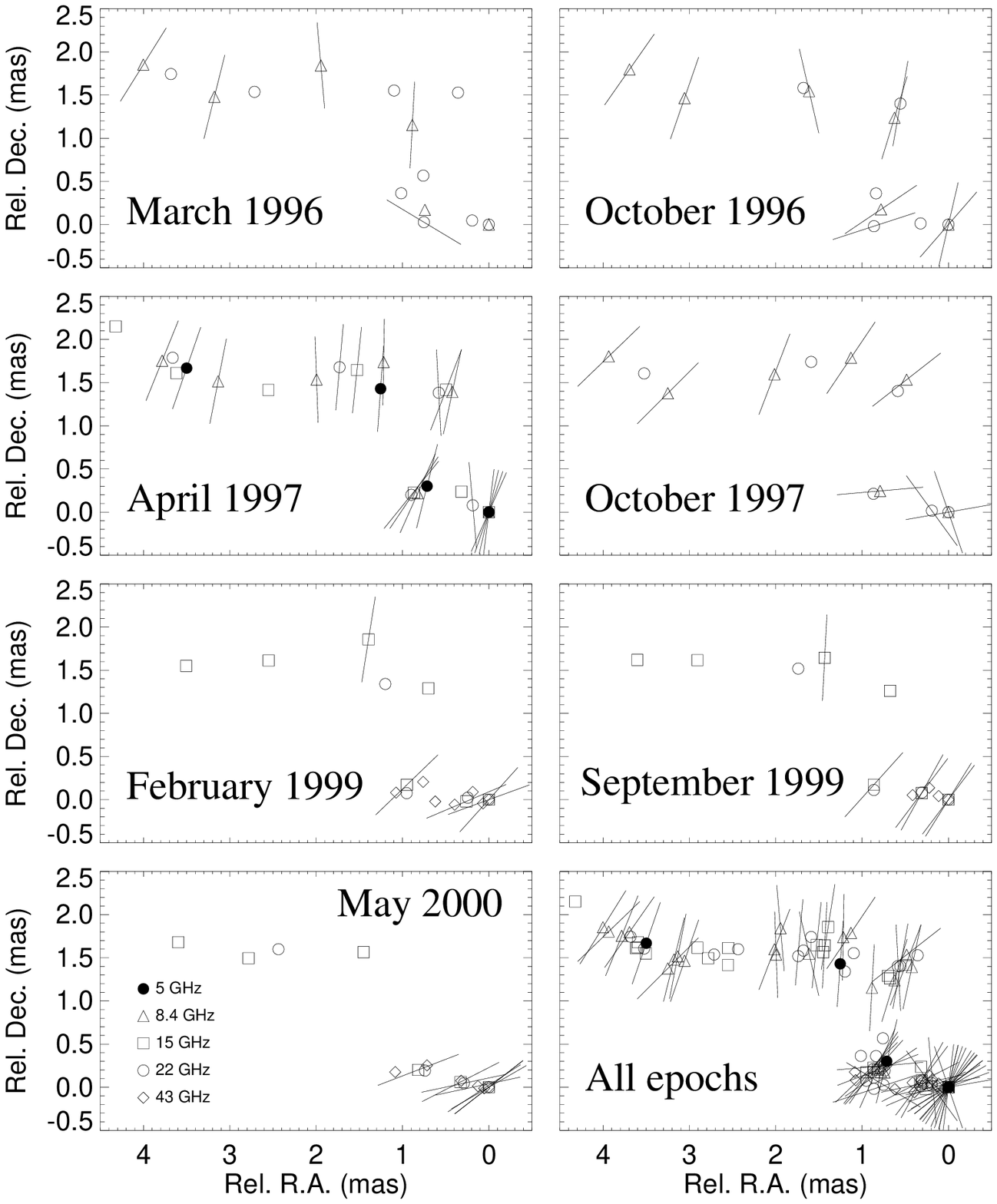}
\caption{Time--dependent $\chi$ distribution at the location
of the model fit components from Tables \ref{fits5} to \ref{fits43}.
The lines represent the orientation of $\chi$, whereas the
observing frequency for each component is shown by a
different symbol (as indicated in the left bottom panel).}
\label{EVPA}
\end{figure}

Fig. \ref{EVPA} represents the $\chi$ values at the locations
of the model fit components in Tables \ref{fits5} to \ref{fits43}.
In agreement with G\'omez et al. (\cite{Gom99}), this
figure shows that the polarization electric vectors were
predominantly perpendicular to the local direction of the northern
jet (outwards from the location of i6) at all the observing epochs.

\begin{figure}
\centering
\includegraphics[bb=0 0 411 604,width=8cm,clip]{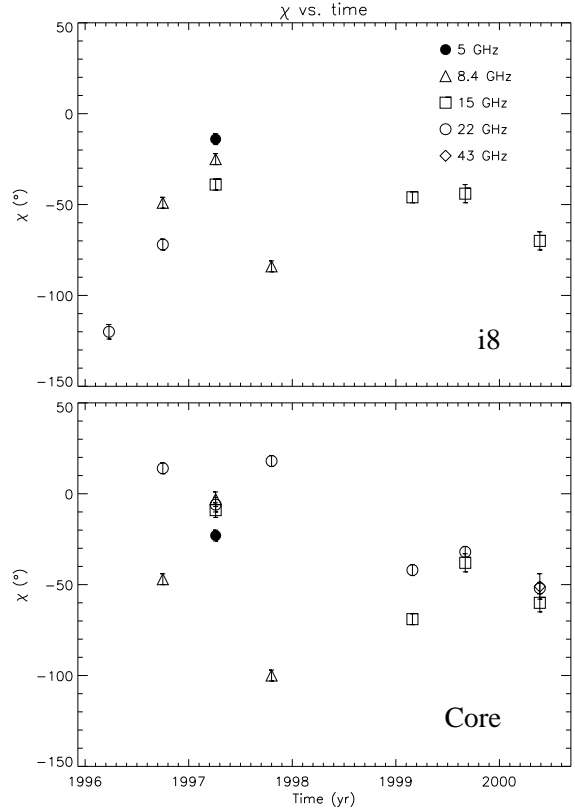}
\caption{$\chi$ as a function of time for i8 (top) and the core
(bottom). The observing frequency for each component is represented
by a different symbol, as indicated in the top plot.}
\label{EVPA_inner}
\end{figure}

For the region between the core and the first strong bend,
$\chi$ shows a clear time dependence, which is more clearly
shown by the plots of $\chi$ as a function of time for i8 and
the core (Fig. \ref{EVPA_inner}).
Note that the $\chi$ behaviour of i9 is very similar to
that of the core.
This is because these two components
belong to the same polarization region, although the model
fits for the total intensity jet are able to separate them.
While the core shows irregular polarization variability from
1996 to 2000 (as is expected from its high level of total
flux density variability), the $\chi$ for i8 linearly increases
at all frequencies between $\sim -120^{\circ}$ in March 1996 to
$\sim -30^{\circ}$ in April 1997.
Furthermore, it experienced an
abrupt rotation by $\sim -60^{\circ}$ in October 1997 and then
maintained a much more stable value of $\sim -45^{\circ}$ during
the last three observing epochs, with a small decrease to
$\sim -25^{\circ}$ observed in May 2000.

The $\sim 90^{\circ}$ monotonic rotation of i8, from 1996 to 1998,
shows evidence of correlation with its total flux density evolution
and also with the integrated total flux density evolution of 0735+178
(see \S \ref{flu}) during the same time range.
This rotation can not be produced by time dependent Faraday rotation,
which would progressively change the relative $\chi$ orientation of
i8 for the different frequencies. A possible explanation for this
behaviour could be either the passage of a shock through
the first bend or the interaction of this bend with a
clumpy external medium, as proposed by Gabuzda, G\'omez \& Agudo
(\cite{GabGA01}), see also \S~\ref{intro}. We favour the jet--external medium
interaction since, although i8 moved outward from the core, it did so with
a slow subluminal speed not typical of moving shocks in relativistic
jets and it displayed a stable position and seemed to be well ``attached''
to the first bend of the underlying jet. In addition, a shock passing
through the first bend would be expected to produce a clockwise rotation
of $\chi$ (for a plane shock perpendicular to the jet axis and moving
following the jet bending towards the north), contrary to what was
shown by our observations.

Note also that the behaviour of i8 is quite different
from that of i6, which maintained a stable $\chi$ orientation
during the time covered by our observations, although its total flux
density monotonically decreased (as for i8).

\section{Summary and conclusions}
\label{concl}

We have investigated the kinematic, total and polarized flux
density evolution of roughly the inner 6\,mas of the VLBI jet
in 0735+178 from 1996 to 2000.

Components i1, i3 and i4
displayed essentially stationary positions during the time spanned
by our observations. They also have stable {\it I} and
{\it P} flux densities and $\chi$ values. This suggests that
the region in which these components are located is mainly composed
of the quiescent underlying jet, whose dynamical time--scales seem
to be longer than those covered by our observations.

For the innermost jet, we have found subluminal or slow
superluminal proper motions with apparent speeds of
($0.58\pm0.10$)\,$c$, ($1.21\pm0.10$)\,$c$ and ($0.65\pm0.53$)\,$c$
for components i8, i9 and i10, respectively.
The variability of the total flux density of the core
allowed us to estimate its Doppler factor $\delta \approx 5$.
Using these measurements and previous estimates of the jet Doppler
factor, we have computed the possible values for the Lorentz
factor $\Gamma$, which ranged from 2 to 4, and the angle to the
line of sight $\theta$ to be $\le 9^\circ$. As we have
not detected fast superluminal motions, such as those
reported for this source before 1993, and the apparent speeds
used to estimate $\theta$ and $\Gamma$ were based on the
slow proper motions of jet features near the jet
bends, our estimates of $\Gamma$ should be related to the
typical pattern speeds of the innermost bent jet structure.
Note that Lorentz factors in bent regions similar to those
estimated for 0735+178 have been shown by three-dimensional
numerical simulations of relativistic helical jets
(Aloy et al.\ \cite{Alo03}).

Therefore, with the exception of i6 (whose position changed
significantly, moving first north and then back south, leading
to a decrease in the curvature of the second bend of the jet),
all the jet components traced during the time range
covered by our monitoring were either quasi-stationary
or showed slow motions, contrary to the situation
observed before 1993 when $7\,c \le \beta_{ap} \le 12\,c$
(Gabuzda et al.~(\cite{Gab94}) and references therein).

A possible explanation for this discrepancy was suggested by
G\'omez et al. (\cite{Gom01}), who proposed a scenario in which
the jet experienced different regimes before 1993 and
after 1995. Before 1993, rapid superluminal motions
($7\,c \le \beta_{ap} \le 12\,c$) were
reported, the structure of the jet remained mainly rectilinear
(Gabuzda et al. \cite{Gab94} and G\'omez et al. \cite{Gom01})
and the radio flux densities from the UMRAO and Mets\"ahovi monitoring
programs were the highest ever detected in 0735+178 (up to
$\sim 5$\,Jy, see Aller et al. \cite{All99} and Ter\"asranta et al.
\cite{Ter04}). This could be called an ``active'' regime of the
jet in 0735+178. From 1995 to 2000, in the ``quiescent'' regime,
only slow motions were detected (see \S \ref{kin}),
the source showed a highly curved structure with two sharp bends
and its integrated flux densities between 5\,GHz and 37\,GHz decreased roughly
linearly up to $\sim 1997$ to $\sim 1998$, reaching its lowest known values
($\aplt 1$\,Jy, Aller et al. \cite{All99} and Ter\"asranta et al. \cite{Ter04}).

This scenario is supported by the 15\,GHz VLBA
images of 0735+178 obtained by the MOJAVE
program\footnote{http://www.physics.purdue.edu/astro/MOJAVE/}
for epochs after November 2002 (Lister \& Homan \cite{Lis05}).
In agreement with the softening of the second bend during
the time covered by our observations, these new images show a fairly
rectilinear jet structure oriented to the north--east (as
reported for early VLBI images before 1993). The typical
double-bend structure displayed by 0735+178 between 1995 and 2000 is
no longer evident.
In addition, the 15\,GHz integrated total flux density
of the source in November 2002 increased by 100\,\% with respect
to our previous observation at 15\,GHz (in May 2000), which seems
to be accompanied by the appearance of rapid superluminal motions
within the innermost four mas of the jet.

In addition, the optical ({\it V}, {\it R} and {\it I} band)
light curves of 0735+178 from January 1995 to March 2001 reported
by Qian \& Tao~(\cite{Qia04}) also agree with the proposed scenario.
The source was observed to decrease its optical flux density from
1995 to January 1998, when it reached one of its faintest levels
ever observed (with a magnitude in {\it V}, $m_{V}=16.68$). After that,
0735+178 became brighter again to reach a new maximum in February 2001
($m_{V}=14.54$). This behaviour is consistent with the one showed
by the radio light curves in Fig.~\ref{lcurv}. Hanski, Takalo
\& Valtaoja~(\cite{Han02}), through the analysis of a different optical
data set from 1988 to 1998 together with the radio light curves from
the Mets\"ahovi monitoring program, showed that the optical flux
density level was always high for 0735+178 when the modelled radio
flux density was at the peak of a flare.

Confirmation of superluminal motions from the data of the MOJAVE
program will provide added support for the proposed bimodal
scenario.
We could therefore consider the possibility of a cyclic behaviour
for 0735+178 in which periods of enhanced activity, with ejection of
superluminal components, are followed by epochs of low activity,
with a highly twisted jet geometry, and vice--versa.
During the high activity period the ejection of superluminal
components, associated with strong shocks, would lead to an
increase of the broad band emission of the source,
enhanced by the large Doppler boosting factors
expected for this source given its small viewing angle.
An increase in the momentum of the component may lead to a more
rectilinear, quasi--ballistic, motion of the component, and
therefore for the apparent jet geometry.
Subsequent low activity periods could reveal the quiescent jet,
characterized by a highly twisted geometry.
This would most probably require a jet with a changing direction
of ejection, since otherwise we would expect the quiescent jet to
follow the straight funnel left by the previous period of high activity.


\begin{acknowledgements}

We gratefully acknowledge Harri Ter\"asranta for providing data
from the Mets\"ahovi monitoring of AGN in advance of
publication.
We gratefully acknowledge Margo F. Aller and Hugh D. Aller for
providing data from the University of Michigan Radio Astronomy
Observatory, which is supported by funds of the University of
Michigan and Barry Clark for providing {\it ad hoc} VLA
time to determine the absolute $\chi$ of the data.
We thank Alan~L. Roy for his helpful suggestions
and Esther S\'anchez for preparing the figures
and tables.
We are also grateful to the anonymous referee for his/her useful
comments.
I. Agudo acknowledges financial support from the EU Commission
through HPRN-CT-2002-00321 project (ENIGMA network).
This research has been supported by the Spanish Ministerio de
Educaci\'on y Ciencia and the European Fund for Regional
Development through grant AYA2004-08067-C03-03.
The work at Boston University was supported by National Science
Foundation grant AST-0406865.

\end{acknowledgements}


\begin{thebibliography}{}

 \bibitem[2001]{Agu01}Agudo, I., G\'omez, J.~L., Mart\'{\i}, J.~M.,
 et al. 2001, ApJ, 549, L183

 \bibitem[1999]{All99}Aller, M.~F., Aller, H.~D., Hughes, P.~A.,
 et al. 1999, ApJ, 512, 601

 \bibitem[2003]{Alo03}Aloy, M.~A., Mart\'{\i}, J.~M., G\'omez, J.~L.,
 et al. 2003, ApJ, 585, L109

 \bibitem[1974]{Car74}Carswell, R.~F., Strittmatter, P.~A., Williams,
 R.~D., et al. 1974, ApJ, 190, L101

 \bibitem[1981]{Fom81}Fomalont, E. 1981, newsletter. NRAO, 3, 3

 \bibitem[1994]{Gab94}Gabuzda, D.~C., Wardle, J.~F.~C., Roberts, D.~H.,
 et al. 1994, ApJ, 435, 128

 \bibitem[2001]{GabGA01}Gabuzda, D.~C., G\'omez, J.~L., Agudo, I.,
 2001 MNRAS, 328, 719

 \bibitem[1999]{Gom99}G\'omez, J.~L., Marscher, A.~P., Alberdi, A.,
 et al. 1999, ApJ, 519, 642

 \bibitem[2000]{Gom00}G\'omez, J.~L., Marscher, A.~P., Alberdi, A.,
 et al. 2000, Science, 289, 2317

 \bibitem[2001]{Gom01}G\'omez, J.~L., Guirado, J.~C., Agudo, I.,
 et al. 2001, MNRAS, 328, 873

 \bibitem[1996]{Gui96}G\"uijosa, A. \& Daly, R.~A. 1996, ApJ, 461, 600

 \bibitem[2002]{Han02} Hanski, M.~T., Takalo, L.~O., Valtaoja, E.
 2002, A\&A, 394, 17

 \bibitem[2001]{Hom01}Homan, D.~C., Ojha, R., Wardle, J.~F.~C.,
 et al. 2001, ApJ, 549, 840

 \bibitem[2005]{Jor05}Jorstad, S.~G., Marscher, A.~P., Lister, M.~L
 et al. 2005, AJ, 130, 1418

 \bibitem[1998]{Kel98}Kellermann, K.~I., Vermeulen, R.~C., Zensus, J.~A.
 et al. 1998, AJ, 115, 1295

 \bibitem[2004]{Kel04}Kellermann, K.~I., Lister, M.~L., Homan, D.~C.
 et al. 2004, ApJ, 609, 539

 \bibitem[1992]{Kol92}Kollgaard, R.~I., Wardle, J.~F.~C., Roberts, D.~H.,
 et al. 1992, AJ, 104, 1687

 \bibitem[1999]{Lah99}L\"ahteenm\"aki, A. \& Valtaoja, E.
  1999, ApJ, 521, 493

 \bibitem[1995]{Lep95}Lepp\"anen, K.~J., Zensus, J.~A. \& Diamond, P.~J.
 1995, AJ, 110, 2479

 \bibitem[2005]{Lis05}Lister, M.~L. \& Homan, D.~C. 2005, AJ, 130, 1389

 \bibitem[2000]{Lov00}Lovell, J., 2000. In Proceedings of the
 VSOP Symposium, Astrophysical Phenomena Revealed by Space VLBI,
 eds. Hirobayashi, H., Eduards, P.G., Murphy, D.W., 301

 \bibitem[2004]{Ojh04}Ojha, R., Homan, D.~C., Roberts, D.~H.,
 et al. 2004, ApJS, 150, 187

 \bibitem[1987]{Pea87}Pearson, T.~J \& Zensus, J.~A. 1987. In Superluminal
 Radio Sources, ed. Zensus, J.~A. \& Pearson, T. (Cambridge Univ. Press), 1

 \bibitem[1994]{Pea94}Pearson, T.~J., Shepherd, M.~C., Taylor G.~B.,
 et al. 1994, AAS, 185, 808

 \bibitem[2004]{Qia04} Qian, B. \& Tao, J. 2004, PASP, 116, 161

 \bibitem[2001]{Rec01}Rector, T.~A. \& Stocke, J.~D. 2001, ApJ, 122, 565

 \bibitem[2004]{Ter04}Ter\"asranta, H., Achren, J., Hanski, M.,
 et al. 2004, A\&A, 427, 769

 \bibitem[1983]{Ulv83}Ulvestad, J.~S., Johnston, K.~J. \& Weiler, K.~W.
 1983, ApJ, 266, 18


\end{thebibliography}
\end{document}